\documentclass[12pt]{article}
\usepackage{amssymb,amsmath,amsthm,amsfonts,amscd}
\newcommand\be{\begin{equation}}
\newcommand\ee{\end{equation}}
\newcommand\bea{\begin{eqnarray}}
\newcommand\eea{\end{eqnarray}}

\newcommand{\fatalpha}{{\bf \alpha \kern -0.44em \alpha}}
\newcommand{\fatsigma}{{\bf \sigma \kern -0.54em \sigma}}
\newcommand{\tpchi}{{\bf \chi \kern -0.35em \chi}}
\newcommand{\llambda}{{\bf \lambda \kern -0.45em \lambda}}

\bibliography{plain}
\title{\bf Entanglement in  multi-qubit  pure fermionic coherent states} %\vspace{20mm}
\title{\bf Entanglement in  multi-qubit  pure fermionic coherent states} %\vspace{20mm}
\author{ G. Najarbashi $^{a}$
 \thanks{E-mail: najarbashi@uma.ac.ir} ,
Y. Maleki $^{a}$
 \\ $^{a}${\small Department of Physics, University of Mohaghegh Ardabili,
Ardabil 179, Iran.}}
\begin{document}
\maketitle
\newpage %\vspace{15mm}

\begin{abstract}

In this paper we investigate the entanglement of multi-qubit
fermionic coherent states  described by anticommutative Grassmann
numbers. Choosing an appropriate weight function, we show that it is
possible to construct some entangled pure states, consisting of
{\bf{GHZ}}, {\bf{W}}, Bell and biseparable  states, by tensor
product of fermion coherent states. Moreover a comparison with
maximal entangled bosonic coherent states is presented and it is
shown that in some cases they have  fermionic counterpart which are
maximal entangled after integration with suitable weight functions.

{\bf PACs Index: 03.65.Ud }
\end{abstract}
\pagebreak

\vspace{7cm}
\section{Introduction}
Understanding the entanglement properties of fermionic systems
remains one of the main goals of quantum information theory
\cite{Nielsen1,Banuls1}. As fermionic optics research grows
\cite{Signal1}, fermionic coherent state (FCS) which is defined as
eigenstate of the fermionic annihilation operator with
Grassman-valued eigenvalue, becomes important. Fermionic coherent
states can be introduced by parametrizing with Grassmann numbers
rather than complex numbers, which overcomes challenges due to
anticommutativity relations \cite{Berezin1,Berezin2,Glauber}.
\par
Besides the idea of constructing bosonic or fermionic coherent
states, there is a great deal of interest in studying entanglement
of them. For bosonic case, some attempts have been made to quantify
the entanglement of multipartite coherent states
\cite{Enk1,Enk2,Fujii1,Wang1,Wang2,Wang3,Wang4,Wang5,Munhoz1,Munro1,Kim1}.
The required conditions for the maximally entangled nonorthogonal
states have been explicitly investigated and some maximally
entangled coherent states  have been classified \cite{Wang1}. But,
despite these attempts, not much work has been done for
investigating the entangled fermionic coherent states
\cite{Khanna1}. The problem is that, strictly within in the
framework of fermion fields, Grassmann numbers which arise  from
Pauli's exclusion principle, anticommute with each other.
\par
The aim of our present work is to elucidate a connection between
fermionic coherent states and multi-qubit pure entangled states,
treated almost always separately. We find throughout this work that
it is quite possible to construct multi-qubit pure entangled states
after integration over tensor product of FCS with suitable weight
function. In particular, for example, we can construct a family of
maximally entangled states like {\bf{GHZ}}, {\bf{W}}, Bell and
Bell-like states \cite{Lewen1} for three qubit systems and then
generalize to the multi-qubit cases. There exist also FCSs which
yield biseparable states in multi-qubits systems except for these
cases we have to consider FCS with different Grassmann numbers.
Also, we make a comparison between maximal entangled bosonic  and
 coherent states, with one complex or Grassmann number in
entries respectively. Somewhat surprisingly, it is shown that in
some particular cases the MESs for fermions, have maximal
counterpart for bosonic coherent states obtained in Ref.
\cite{Wang1} via  concurrence measure \cite{Wootters1}.
\par
The paper is organized as follows. In section $2$, the FCS for two
level system is introduced. In section $3$, explicit examples of
multi-qubit entangled states such as Bell, Bell-like, $\mathbf{W}$,
$\mathbf{GHZ}$ and biseparable states are constructed by tensor
product of FCS with appropriate weight functions. Section $4$ is
devoted to compare some special MESs of bosonic and fermionic
coherent states  with just one complex or Grassman number in each
entries. The paper ends with a brief conclusion.
\section{ Grassmannian Coherent States}
For our purpose, it is necessary to study the mathematical structure
of the anti-commuting mathematical objects, so called Grassmann
algebra which are needed in order to construct relevant coherent
states \cite{Berezin1,Berezin2,Glauber}. To describe this algebra we
consider n generators $ \{\theta_{1},...,\theta_{n}\}$ satisfying
the relations:
\begin {equation}\label{comut}
\{\theta_{i} ,\theta_{j}\} = 0\quad \forall\ i, j=1,2,...,n,
\end {equation}
and clearly we have
\begin {equation}\label{nilpot}
\theta_{i}^2=0\quad \forall i=1,2,...,n.
\end {equation}
 Any linear combination of ${\theta_{i}}$ with the complex
number coefficients is called Grassmann number. In other words, we
consider Taylor expansion of a Grassmann function as follows
$$
g(\theta_{1},\theta_{2},...\theta_{n})=c_{0}+\sum_{i}c_{i}\theta_{i}+\sum_{i<j}c_{i,j}\theta_{i}\theta_{j}+...,
$$
where $c_{0},c_{i},c_{i,j},...\in \mathbb{C}$. For instance, $\exp
(\theta_{1}\theta_{2})=1+\theta_{1}\theta_{2}$. The complex
conjugate of the Grassmann number $\theta$ is also defined by $
(\theta)^*= \theta^*,$ which is treated as another Grassmann number.
A Grassmann function is called "of $n$ degree" if it contains a term
with $n$ multiple of Grassmann numbers like
$\theta_{1}\theta_{2}...\theta_{n}$. The Grassmann integration and
differentiation over the complex Grassmann variables are given by
\begin {equation}
\int d\theta f(\theta)= \frac{\partial f(\theta)}{\partial \theta}
\end {equation}
\begin {equation}
   \int d\theta=0,  \quad \int d\theta \theta=1,\quad  \int
 d\theta^*=0, \quad \int d\theta^*\theta^*=\theta^*,
\end {equation}
\begin {equation}
\frac{\partial}{\partial\theta}\theta=1,\quad  \frac{\partial}
{\partial \theta^*}1=0,\quad
   \frac{\partial} {\partial \theta}1=0,\quad  \frac{\partial}{\partial
\theta^*}\theta^*=1
\end {equation}
\begin {equation}
\frac{\partial^2}{\partial^2\theta}=\frac{\partial^2}{\partial^2\theta^*}=0.
\end {equation}
For the next uses, we require the following quantization relations
between Fock states and Grassmann numbers
\begin {equation}
\begin{array}{c}
  \theta|0\rangle=|0\rangle \theta,\qquad  \theta|1\rangle=-|1\rangle
\theta, \\
  \theta\langle0|=\langle0| \theta,\qquad
  \theta\langle1|=-\langle1|,
\theta
\end{array}
\end {equation}
which implies that Grassmann numbers commute with
$|0\rangle\langle0|,$ and $|1\rangle\langle1|,$ while anticommute
with $|1\rangle\langle0|,$ and $|0\rangle\langle1|.$ Now let $a$ and
${a}^\dag$ be annihilation and creation operators for a fermionic
system respectively. These operators satisfy the anti-commutation
relations
\begin {equation}
\begin{array}{c}
  \{a,{a}^\dag \}=1, \\
  \{a,{a}\}=\{{a}^\dag,{a}^\dag \}=0.
\end{array}
\end {equation}
Clearly, $a$ and ${a}^\dag$ are nilpotent. We shall also assume that
Grassmann variables anti-commute with fermionic operators
\begin {equation}
\{a,\theta\}=\{a^\dag,\theta\}=0.
\end {equation}
A fermionic coherent state, like the bosonic case, is defined as
eigen-state of the annihilation operator
\begin {equation}\label{coherent1}
a|\theta\rangle=\theta|\theta\rangle,
\end {equation}
which is satisfied by the following state
\begin {equation}\label{coherent2}
|\theta\rangle=\exp(\frac{-\theta^*\theta}{2})
(|0\rangle-\theta|1\rangle)=D(\theta)|0\rangle.
\end {equation}
where
\begin {equation}
D(\theta):=\exp(a^\dag\theta-\theta^* a)
\end {equation}
Note that the displacement operator $D(\theta)$ is a unitary
operator i.e; $D(\theta)D(\theta)^\dag=I$.
\section{Entanglement and FCS}
In this section, we show that one can get the well known maximally
entangled pure states such as {\bf{GHZ}}, {\bf{W}} , Bell and
Bell-like states \cite{Lewen1}, through integrating over tensor
product of FCSs with suitable choice of weight function.
\subsection{Bell and Bell-like states}
 Let us consider the simple cases that yield the following  Bell states
\begin {equation}\label{bell1}
|{\Psi}^{\pm}\rangle=\frac{1}{\sqrt{2}}(|01\rangle\pm|10\rangle),
\end {equation}
\begin {equation}\label{bell2}
|{\Phi}^{\pm}\rangle=\frac{1}{\sqrt{2}}(|00\rangle\pm|11\rangle).
\end {equation}
Regarding the FCS of Eq. (\ref{coherent2}), we product two  states
$|\theta\rangle $  and $|\pm\theta\rangle $ as follows
\begin {equation}
|\theta\rangle |\pm\theta\rangle=\exp(-\theta^*\theta)[ |00\rangle
\mp\theta|01\rangle-\theta|10\rangle],
\end {equation}
\begin {equation}\label{coherent3}
|-\theta\rangle |\mp\theta\rangle=\exp(-\theta^*\theta)[ |00\rangle
\pm\theta|01\rangle+\theta|10\rangle],
\end {equation}
and by subtracting the above states we get
\begin {equation}
|\theta\rangle |\pm\theta\rangle-|-\theta\rangle
|\mp\theta\rangle=\mp2\theta(|01\rangle\pm|10\rangle).
\end {equation}
Our task is to find the weight function such that when we integrate
over Grassmann numbers, $\theta$ and $\theta^*$ yield the
$|{\Psi}^{(\pm)}\rangle$. To this aim let
\begin {equation}
w(\theta,\theta^*)=c_{0}+c_{1}\theta+c_{2}\theta^*+c_{3}\theta^*\theta,
\end {equation}
be such weight function. Then
$$
\int d\theta^* d\theta
\ w (\theta,\theta^*)[|\theta\rangle
|\pm\theta\rangle-|-\theta\rangle |\mp\theta\rangle]=
|{\Psi}^{\pm}\rangle,
$$
where it is satisfied with $c_{2}= \pm\frac{1}{2\sqrt{2}}$, and the
other coefficients  are arbitrary which may be taken as
$c_{0}=c_{1}=c_{3}=0$ i.e.,
$w(\theta,\theta^*)=\pm\frac{1}{2\sqrt{2}}\theta^*$. Note that
instead of subtracting the states $|\theta\rangle |\pm\theta\rangle$
and $|-\theta\rangle |\mp\theta\rangle$, we can add them and
integrate on  weight function $w (\theta,\theta^*)=\pm\frac{1}{2}$,
which in turn yields the separable state $|00\rangle$. Another
tensor product of FCS which yields the Bell state
$|{\Psi}^{-}\rangle$ are
$$
\int d\theta^* d\theta \ w (\theta,\theta^*)|\pm\theta\rangle
|\pm\theta\rangle= |{\Psi}^{-}\rangle, \qquad \mathrm{with}\qquad
w(\theta,\theta^*)=\frac{\pm1}{\sqrt{2}}\theta^*.
$$
This is not the only way to construct $|{\Psi}^{(\pm)}\rangle$
states by FCS. We can get the same result if we define the states
\begin {equation}\label{GCS pm}
|{\theta}\rangle_{\pm}=|\theta\rangle\pm|-\theta\rangle,
\end {equation}
and by integration as follows we have
\begin {equation}\label{addeven}
\int d\theta^* d\theta\ w (\theta,\theta^*)\
[|{\theta}\rangle_+|{\theta}\rangle_-\pm|{\theta}\rangle_-|{\theta}\rangle_+]=|{\Psi}^{\pm}\rangle,
\end {equation}
where
$$
 w(\theta,\theta^*)=\frac{1}{4\sqrt{2}}\theta^*.
$$
It is interesting that we can construct all Bell states if we assume
the FCS to be constructed by two Grassmann numbers $\theta_{1}$ and
$\theta_{2}$ and their complex conjugations. Such a possible state
may be $|\theta\rangle |\theta^*\rangle$, whose integration with
weight function $ w(\theta,\theta^*)=\exp(\pm\theta\theta^*)$, gives
\begin {equation}
\int d\theta^* d\theta (\frac{\pm1}{\sqrt{2}}\
e^{\pm\theta\theta^*}\ )|\theta^*\rangle|\theta\rangle
=|{\Phi}^{\pm}\rangle,
\end {equation}
And also
\begin {equation}
\int d\theta^* d\theta \frac{1}{\sqrt{2}}(\theta^*\pm\theta)
|\theta^*\rangle|\theta\rangle =|{\Psi}^{\pm}\rangle.
\end {equation}
To generalize more we take some other states, the first of which,
goes as follows
\begin {equation}
|\theta_{1}\rangle|\theta_{2}\rangle=\exp[\frac{-1}{2}(\theta_{1}^*\theta_{1}+\theta_{2}^*\theta_{2})]
(|00\rangle-\theta_{2}|01\rangle-\theta_{1}|10\rangle-\theta_{1}\theta_{2}|11\rangle)
\end {equation}
The above state leads to Bell states via suitable weight functions.
For example
\begin {equation}
\int d\theta_{1}^* d\theta_{1} d\theta_{2}^* d\theta_{2}
\frac{1}{\sqrt{2}}
(\theta_{1}^*\theta_{1}\theta_{2}^*\theta_{2}\mp\theta_{1}^*\theta_{2}^*)
|\theta_{1}\rangle|\theta_{2}\rangle=|{\Phi}^{\pm}\rangle,
\end {equation}
\begin {equation}
\int d\theta_{1}^* d\theta_{1} d\theta_{2}^* d\theta_{2}
(\frac{-1}{\sqrt{2}})(\theta_{1}^*\theta_{1}\theta_{2}^*+\theta_{1}^*\theta_{2}^*\theta_{2})
|\theta_{1}\rangle|\theta_{2}\rangle=|{\Psi}^{\pm}\rangle.
\end {equation}
We note that in the case $|\theta\rangle |\theta^*\rangle$, it is
impossible to choose a weight function of degree three or more,
while in the case $|\theta_{1}\rangle|\theta_{2}\rangle$, it is
possible. Now consider the symmetric and anti-symmetric FCSs
$$
\hspace{-5cm}|\Lambda_{\pm}(\theta_{1},\theta_{2})\rangle=|\theta_{1}\rangle|\theta_{2}\rangle\pm|\theta_{2}\rangle|\theta_{1}\rangle=
\pm|\Lambda_{\pm}(\theta_{2},\theta_{1})\rangle,
$$
which the following maximal entangled and separable states are
deduced
\begin {equation}
\int d\theta_{1}^* d\theta_{1} d\theta_{2}^*
d\theta_{2}\big(\frac{2}{\sqrt{2}}\theta_{1}^*\big)\
|\Lambda_{\pm}(\theta_{1},\theta_{2})\rangle= |\Psi^{(\pm)}\rangle,
\end {equation}
\begin {equation}
\int d\theta_{1}^* d\theta_{1} d\theta_{2}^*
d\theta_{2}w(\theta_{1},\theta_{1}^*,\theta_{2},\theta_{2}^*)
|\Lambda_{\pm}(\theta_{1},\theta_{2})\rangle=\left\{\begin{array}{c}
                                                      \hspace{-.8cm}|00\rangle \quad \mathrm{with}\quad w=\frac{1}{2}\theta_{1}^*\theta_{2}^*\\
                                                     |11\rangle  \quad \mathrm{with} \quad
                                                     w=\frac{1}{2}\theta_{1}^*\theta_{1}\theta_{2}^*\theta_{2}.
                                                    \end{array}
\right.
\end {equation}
The anti-symmetric  state
 $|\Lambda_{-}(\theta_{1},\theta_{2})\rangle $
only gives Bell state $ |{\Psi}^{-}\rangle$ which is anti-symmetric,
and the symmetric  state
 $|\Lambda_{+}(\theta_{1},\theta_{2})\rangle $
only gives Bell state $ |{\Psi}^{+}\rangle$ which is symmetric.
\par
Another MESs which can be manipulated by FCSs are Bell-like states
\begin {equation}\label{Bell-like1}
|{\Psi}^{\pm}\rangle_{BL}=\frac{1}{\sqrt{2}}({e^{
i\frac{\pi}{4}}|01\rangle\pm e^{-i\frac{\pi}{4}}|10\rangle}).
\end {equation}
where
 \begin {equation}\label{Bell-like2}
\int d\theta^* d\theta\ \frac{1}{\sqrt{2}}\left({e^{
i\frac{\pi}{4}}\ \theta^*\pm e^{-i\frac{\pi}{4}}\ \theta}\right)
|\theta^*\rangle|\theta\rangle=|{\Psi}^{\pm}\rangle_{BL},
\end {equation}
\begin {equation}
\int d\theta_{1}^* d\theta_{1} d\theta_{2}^* d\theta_{2}
\frac{1}{\sqrt{2}} \left(e^{ i\frac{\pi}{4}}\
\theta_{1}\theta_{1}^*\theta_{2}^*\pm e^{-i\frac{\pi}{4}}\
\theta_{1}^*\theta_{2}\theta_{2}^* \right)
|\theta_{1}\rangle|\theta_{2}\rangle=|{\Psi}^{\pm}\rangle_{BL}.
\end {equation}
\subsection{GHZ and W states }
Here, we proceed the same way as above to construct three qubit
MESs, known as $\mathbf{W}$ and $\mathbf{GHZ}$ states which are used
widely in quantum information theory. Then we generalize them for
n-qubit cases. For $\mathbf{W}$ case, consider tensor product of
three FCSs of the form
$$
|\theta\rangle|\theta\rangle|\theta\rangle=\exp-\frac{3}{2}\theta^*\theta(|000\rangle-\theta(|001\rangle+|010\rangle+|100\rangle)),
$$
thus with a convenient weight function we get
\begin {equation}
\int
d\theta^*d\theta(\frac{\theta^*}{\sqrt{3}})(|\theta\rangle|\theta\rangle|\theta\rangle)=
\frac{1}{\sqrt{3}}(|100\rangle+|010\rangle+|001\rangle).
\end {equation}
One can easily generalize this to n-qubit $\mathbf{W}$ state as
follows
\begin {equation}
\int
d\theta^*d\theta(\frac{\theta^*}{\sqrt{n}})\underbrace{|\theta\rangle|\theta\rangle...|\theta\rangle}_{n\
times} =|{\bf{W}}^{(n)}\rangle,
\end {equation}
where
\begin {equation}
|{\bf{W}}^{(n)}\rangle=\frac{1}{\sqrt{n}}(|100...0\rangle+|010...0\rangle+
...+|0...001\rangle).
\end {equation}
It is convenient to write the n-qubit $\mathbf{W}$ states with
respect to FCSs of Eq.(\ref{GCS pm}) as
\begin {equation}
\int d\theta^*d\theta(\frac{1}{2^n\sqrt{n}}\theta^*)|\psi\rangle
=|{\bf{W}}^{(n)}\rangle,
\end {equation}
where
\begin {equation}
|\psi\rangle=|{\theta}\rangle_{+}|{\theta}\rangle_{+}...|{\theta}\rangle_{+}|{\theta}\rangle_{-}+
|{\theta}\rangle_{+}|{\theta}\rangle_{+}...|{\theta}\rangle_{-}|{\theta}\rangle_{+}+
...+|{\theta}\rangle_{-}|{\theta}\rangle_{+}...|{\theta}\rangle_{+}|{\theta}\rangle_{+}.
\end {equation}
To construct the three qubit $\mathbf{GHZ}$ state we have to use
tensor product of three FCSs with different Grassmann numbers
$|\theta_{1}\rangle|\theta_{2}\rangle|\theta_{3}\rangle$. Then, the
integration goes as
\begin {equation}
\int
d\theta_{1}^*d\theta_{1}d\theta_{2}^*d\theta_{2}d\theta_{3}^*d\theta_{3}
[\frac{1}{\sqrt{2}}(\theta_{1}^*\theta_{1}\theta_{2}^*\theta_{2}\theta_{3}^*\theta_{3}+\theta_{1}^*\theta_{2}^*\theta_{3}^*)]
|\theta_{1}\rangle|\theta_{2}\rangle|\theta_{3}\rangle=\frac{1}{\sqrt{2}}
(|000\rangle+|111\rangle).
\end {equation}
 In a similar way as n-qubit $\mathbf{W}$ states, we can create the general n-qubit  $\mathbf{GHZ}$
states by using FCSs. To this aim we take
$|\theta_{1}\rangle|\theta_{2}\rangle...|\theta_{n}\rangle$ together
with  weight function as
\begin {equation}
w=\frac{1}{\sqrt{2}}(\theta_{1}^*\theta_{1}\theta_{2}^*\theta_{2}...\theta_{n}^*\theta_{n}+\theta_{1}^*\theta_{2}^*...\theta_{n}^*),
\end {equation}
 we get
\begin {equation}
\int
d\theta_{1}^*d\theta_{1}d\theta_{2}^*d\theta_{2}...d\theta_{n}^*d\theta_{n}w|\theta_{1}\rangle|\theta_{2}\rangle...|\theta_{n}\rangle=\frac{1}{\sqrt{2}}
(|00...0\rangle+|11...1\rangle)=|{\bf{GHZ}}^{(n)}\rangle.
\end {equation}
\subsection{Biseparability}
Here, we use FCSs to obtain biseparabile states which, depending on
how one considers partition for given state, there exists an
entanglement in their subsystems \emph{partially}. For example if a
pure state $|\psi\rangle_{ABC}$ involves the three subsystems $A,B$
and $C$, the partition $\{A\}$ may be separable while $\{B,C\}$ are
entangled. As an illustration, let us consider three and four
partite cases as some examples. Hence, the entanglement of bipartite
states can be made by FCS
$|\theta_{1}\rangle|\theta_{2}\rangle|\theta_{3}\rangle$
 as follows
\begin {equation}
\int
d\theta_{1}^*d\theta_{1}d\theta_{2}^*d\theta_{2}d\theta_{3}^*d\theta_{3}
[\frac{1}{\sqrt{2}}(\theta_{1}^*\theta_{1}\theta_{2}^*\theta_{3}\theta_{3}^*\pm\theta_{1}^*\theta_{1}\theta_{2}\theta_{2}^*\theta_{3}^*)]
|\theta_{1}\rangle|\theta_{2}\rangle|\theta_{3}\rangle=
 |0\rangle_{1}|{\Psi}^{\pm}\rangle_{2,3}.
\end {equation}
where it implies that the entanglement is just between the second
and third qubits. Other biseparable states are
\begin {equation}
\int
d\theta_{1}^*d\theta_{1}d\theta_{2}^*d\theta_{2}d\theta_{3}^*d\theta_{3}
[\frac{1}{\sqrt{2}}(\theta_{1}\theta_{1}^*\theta_{2}\theta_{2}^*\theta_{3}\theta_{3}^*\pm\theta_{1}^*\theta_{1}\theta_{2}^*\theta_{3}\theta_{3}^*)]
|\theta_{1}\rangle|\theta_{2}\rangle|\theta_{3}\rangle=
 |{\Psi}^{\pm}\rangle_{1,2}|0\rangle_{3},
\end {equation}
\begin {equation}
\int
d\theta_{1}^*d\theta_{1}d\theta_{2}^*d\theta_{2}d\theta_{3}^*d\theta_{3}
[\frac{1}{\sqrt{2}}(\theta_{1}^*\theta_{2}\theta_{2}^*\theta_{3}\theta_{3}^*\pm\theta_{1}^*\theta_{1}\theta_{2}\theta_{2}^*\theta_{3}^*)]
|\theta_{1}\rangle|\theta_{2}\rangle|\theta_{3}\rangle=
 |0\rangle_{2}|{\Psi}^{\pm}\rangle_{1,3}
\end {equation}
where
$$
|0\rangle_{2}|{\Psi}^{\pm}\rangle_{1,3}=\frac{1}{\sqrt{2}}(|001\rangle\pm|100\rangle)
$$
Furthermore one can easily see that
\begin {equation}
\int
d\theta_{1}^*d\theta_{1}d\theta_{2}^*d\theta_{2}d\theta_{3}^*d\theta_{3}
[\frac{1}{\sqrt{2}}(\theta_{1}\theta_{1}^*\theta_{2}\theta_{2}^*\theta_{3}\theta_{3}^*\pm\theta_{1}\theta_{1}^*\theta_{3}^*\theta_{2}^*)]
|\theta_{1}\rangle|\theta_{2}\rangle|\theta_{3}\rangle=
 |0\rangle_{1}|{\Phi}^{\pm}\rangle_{2,3}.
\end {equation}
The biseparable states $|0\rangle_{2}|{\Phi}^{\pm}\rangle_{1,3}$ and
$|0\rangle_{3}|{\Phi}^{\pm}\rangle_{1,2}$ can be obtained in a same
manner as above with different weight functions. We can also
construct four qubit biseparabile states like the three qubit case.
To do this we take
$|\theta_{1}\rangle|\theta_{2}\rangle|\theta_{3}\rangle|\theta_{4}\rangle
$, and choose a weight function as
\begin {equation}
w=\frac{1}{\sqrt{3}}(\theta_{1}\theta_{1}^*\theta_{2}^*\theta_{3}\theta_{3}^*\theta_{4}\theta_{4}^*+
\theta_{1}\theta_{1}^*\theta_{2}\theta_{2}^*\theta_{3}^*\theta_{4}\theta_{4}^*+
\theta_{1}\theta_{1}^*\theta_{2}\theta_{2}^*\theta_{3}\theta_{3}^*\theta_{4}^*),
\end {equation}
then we have
\begin {equation}\label{Wbisep}
\int
d\theta_{1}^*d\theta_{1}d\theta_{2}^*d\theta_{2}d\theta_{3}^*d\theta_{3}d\theta_{4}^*d\theta_{4}
w|\theta_{1}\rangle|\theta_{2}\rangle|\theta_{3}\rangle|\theta_{4}\rangle=
 |0\rangle_{1}|{{\bf{W}}^{(3)}}\rangle_{2,3,4},
\end {equation}
where it means that the first qubit is not entangled with the other
three qubit related to partition $\{2,3,4\}$. One can obtain
biseparable states $ |s\rangle_{i}|{{\bf{W}}^{(3)}}\rangle_{j,k,l},$
  $(s=0,1)$ which may be any partition
  as Eq.(\ref{Wbisep}). From both the partition and type of
  entanglement point of view, there are some other possibilities such as
\begin {equation}
\int
d\theta_{1}^*d\theta_{1}d\theta_{2}^*d\theta_{2}d\theta_{3}^*d\theta_{3}d\theta_{4}^*d\theta_{4}
w|\theta_{1}\rangle|\theta_{2}\rangle|\theta_{3}\rangle|\theta_{4}\rangle=
 |0\rangle_{1}|{{\bf{GHZ}}^{(3)}}\rangle_{2,3,4},
\end {equation}
where
$$
w=\frac{1}{\sqrt{2}}(\theta_{1}\theta_{1}^*\theta_{2}\theta_{2}^*\theta_{3}\theta_{3}^*\theta_{4}\theta_{4}^*+
\theta_{1}\theta_{1}^*\theta_{2}^*\theta_{3}^*\theta_{4}^*),
$$
and
\begin {equation}
\int
d\theta_{1}^*d\theta_{1}d\theta_{2}^*d\theta_{2}d\theta_{3}^*d\theta_{3}d\theta_{4}^*d\theta_{4}
w|\theta_{1}\rangle|\theta_{2}\rangle|\theta_{3}\rangle|\theta_{4}\rangle=
 |00\rangle_{1,2}|{\Phi}^{\pm}\rangle_{3,4},
\end {equation}
with
$$
w=\frac{1}{\sqrt{2}}(\theta_{1}\theta_{1}^*\theta_{2}\theta_{2}^*\theta_{3}\theta_{3}^*\theta_{4}\theta_{4}^*\pm
\theta_{1}\theta_{1}^*\theta_{2}\theta_{2}^*\theta_{3}^*\theta_{4}^*),
$$
and also
\begin {equation}
\int
d\theta_{1}^*d\theta_{1}d\theta_{2}^*d\theta_{2}d\theta_{3}^*d\theta_{3}d\theta_{4}^*d\theta_{4}
w|\theta_{1}\rangle|\theta_{2}\rangle|\theta_{3}\rangle|\theta_{4}\rangle=
 |{\Psi}^{+}\rangle_{1,2}|{\Phi}^{+}\rangle_{3,4},
\end {equation}
with
$$
w=\frac{1}{2}(\theta_{1}^*\theta_{2}\theta_{2}^*\theta_{3}\theta_{3}^*\theta_{4}\theta_{4}^*+
\theta_{1}\theta_{1}^*\theta_{2}^*\theta_{3}\theta_{3}^*\theta_{4}\theta_{4}^*+
\theta_{1}^*\theta_{2}\theta_{2}^*\theta_{3}^*\theta_{4}^*+
\theta_{1}\theta_{1}^*\theta_{2}^*\theta_{3}^*\theta_{4}^*).
$$
It is easy to develop this discussion to more general forms.
\section{Comparison with bosonic coherent states}
It is tempting to compare the fermion and boson coherent states. A
bosonic coherent state can be defined as eigen-state of the
annihilation operator
\begin {equation}\label{coherent1}
b|\alpha\rangle=\alpha|\alpha\rangle,
\end {equation}
where $\alpha$ is a complex number, and $b$ is annihilation operator
for the bosonic coherent state
\begin {equation}\label{coherent2}
|\alpha\rangle=e^{\frac{-|\alpha|^2}{2}}
\sum_{n=0}^\infty\frac{\alpha^n}{\sqrt{n!}}|n\rangle=D(\alpha)|0\rangle.
\end {equation}
where the displacement operator $D(\alpha)$ is
\begin {equation}
D(\alpha):=\exp(b^\dag\alpha-\alpha^*b)
\end {equation}
There are different measures to quantify the entanglement of a
quantum system (for a good review see \cite{Plenio1}). One of them
is entanglement of formation which gives the exact formula based on
the often used two-qubit concurrence defined as \cite{Wootters1}
\begin {equation}\label{concurrence}
C=|\langle\zeta|\sigma_y\otimes\sigma_y|\zeta^*\rangle|,
\end {equation}
where $\sigma_y$ is $y$ component of the usual Pauli spin matrices.
The concurrence of the following state
\begin {equation}\label{generalbos}
|\zeta\rangle=\mu|\alpha\rangle|\beta\rangle+\nu|\gamma\rangle|\delta\rangle,
\end {equation}
in the subspace spanned by
$|\alpha\rangle,|\beta\rangle,|\gamma\rangle$ and $|\delta\rangle$
is
\begin {equation}
C=\frac{|\mu\nu|\sqrt{(1-|\langle\alpha|\gamma\rangle|^2)(1-|\langle\beta|\delta\rangle|^2)}}
{|\mu|^2+|\nu|^2+\mu\nu^*\langle\gamma|\alpha\rangle\langle\delta|\beta\rangle+\mu^*\nu\langle\alpha|\gamma\rangle\langle\beta|\delta\rangle},
\end {equation}
where $|\alpha\rangle,|\beta\rangle,|\gamma\rangle$ and
$|\delta\rangle$ are bosonic coherent states. The denominator of the
above concurrence come from the normalization of $|\zeta\rangle$.
When $C=1$ then $|\zeta\rangle$ is MES that is the conditions for
maximality of entanglement for nonorthogonal four bosonic coherent
states in Eq. (\ref{generalbos}) are \cite{Wang1}
\begin {equation}\label{maxcond}
\mu=\nu e^{(i\varphi)}\quad \mathrm{and} \quad
\langle\alpha|\gamma\rangle=-\langle\delta|\beta\rangle
e^{(i\varphi)}.
\end {equation}
For the particular cases we will discuss two following examples
\begin {equation}
|k_{1}\alpha\rangle|k_{2}\alpha\rangle\pm|k_{3}\alpha\rangle|k_{4}\alpha\rangle\quad,
\qquad k_{i}\in \mathbb{C},
\end {equation}
where due to their different  behaviors, under imposing the
maximality conditions, we will treat them separately.
\subsection{Example 1}
Consider the following bosonic coherent state
\begin {equation}\label{bospn genera2}
|k_{1}\alpha\rangle|k_{2}\alpha\rangle-|k_{3}\alpha\rangle|k_{4}\alpha\rangle\quad,
\qquad k_{i}\in \mathbb{C},
\end {equation}
 where its  concurrence goes as
\begin {equation}
C=\frac{2\left[(1-e^{{-\frac{1}{2}|\alpha|^2}(f_{13}+f_{13}^*)})(1-e^{{-\frac{1}{2}|\alpha|^2}(f_{24}+f_{24}^*)})\right]^{1/2}}
{2-e^{{-\frac{1}{2}|\alpha|^2(f_{13}^*+f_{24}^*)}}-e^{{-\frac{1}{2}|\alpha|^2(f_{13}+f_{24})}}},
\end {equation}
where
$$
f_{ij}=|k_i|^2+|k_j|^2-2k_i^*k_{j}\ .
$$
 Regarding the conditions (\ref{maxcond}) we have
$f_{13}=f_{24}^*$, which implies that they have the same real and
imaginary parts i.e.,
\begin {equation}\label{realcon}
|k_{1}-k_{3}|=|k_{2}-k_{4}|\quad,\ \ \mathrm{or} \quad
(k_{1}-k_{3})=(k_{2}-k_{4})e^{i\phi}
\end {equation}
\begin {equation}\label{imagecon}
Im(k_1^*k_3)=Im(k_4^*k_2).
\end {equation}
Some special cases of MES for the state (\ref{bospn genera2}), up to
a normalization factor, are deduced \cite{Wang1}
\begin {equation}\label{boson1}
\hspace{-6cm}|\psi\rangle_{boson}=\left\{\begin{array}{c}
  |\alpha\rangle|-\alpha\rangle-|-\alpha\rangle|-3\alpha\rangle, \\
  \hspace{-.7cm}|\alpha\rangle|-\alpha\rangle-|-\alpha\rangle|\alpha\rangle, \\
  \hspace{-.9cm}|\alpha\rangle|\alpha\rangle-|i\alpha\rangle|-i\alpha\rangle, \\
  \hspace{-.9cm}|\alpha\rangle|-\alpha\rangle-|i\alpha\rangle|i\alpha\rangle.
\end{array}\right.
\end {equation}
Now we return to tensor product of FCSs and consider the same form
of equation (\ref{bospn genera2}) where the complex parameter
$\alpha$ is replaced by Grassmann number $\theta$ as follows
\begin {equation}\label{generalfer}
|k_{1}\theta\rangle|k_{2}\theta\rangle-|k_{3}\theta\rangle|k_{4}\theta\rangle.
\end {equation}
We call a FCS \emph{maximal}, if there is a Grassmann weight
function whose integration over that FCS gives a MES. If we take the
weight function as
\begin {equation}\label{weight1}
w(\theta,\theta^*)=\frac{1}{m\sqrt{2}}\theta^*,
\end {equation}
then
$$
|{\Psi}\rangle_{max}=\int d\theta^* d\theta \ w
(\theta,\theta^*)[|k_{1}\theta\rangle|k_{2}\theta\rangle-|k_{3}\theta\rangle|k_{4}\theta\rangle]
$$
\begin {equation}\label{int-gen}
\hspace{1cm}= \frac{1}{m\sqrt{2}}[k_{2}- k_{4}]|01\rangle+
\frac{1}{m\sqrt{2}}[k_{1}- k_{3}]|10\rangle.
\end {equation}
This state is MES if
\begin {equation}\label{condfer}
(k_{1}-k_{3})=e^{i\phi}(k_{2}-k_{4})=m.
\end {equation}
which is equivalent to (\ref{realcon}). Now we are interested in
finding some especial cases of $|{\Psi}\rangle_{max}$ which lead to
MESs similar to the four bosonic coherent states of the form
$|\psi\rangle_{boson}$. We distinguish the following cases:
\\
 \textbf{Cases 1,2:}\quad
Let $|{\Psi}\rangle_{max}=|{\Psi}^{\pm}\rangle$, $\phi=0$ and
$m=\pm2$. Imposing $k_{1}=-k_{2}=-k_{3}$, the Eq. (\ref{generalfer})
reduces to
\begin {equation}
|\theta\rangle|-\theta\rangle-|-\theta\rangle|-3\theta\rangle,
\end {equation}
\begin {equation}
|\theta\rangle|-\theta\rangle-|-\theta\rangle|\theta\rangle.
\end {equation}
where first FCS refers to plus sign and the second one refers to
minus sign.
\\
\textbf{Cases 3,4:}\quad Let
$|{\Psi}\rangle_{max}=|{\Psi}^{\pm}\rangle_{BL}$, $\phi=0$ and
$m=\pm\sqrt{2}$.
 If we take $k_{1}=1,k_{2}=\pm1,k_{3}=k_{4}=\mp i$, then FCS
 (\ref{generalfer}) is reduced to the following states
\begin {equation}
|\theta\rangle |\pm\theta\rangle-|i\theta\rangle|\mp i\theta\rangle
\end {equation}
The above FCSs obtained in cases 1-4 could be compared with  the
maximal bosonic coherent states $|\psi\rangle_{boson}$. Of course,
we deliberately call these FCSs maximally entangled as done for
bosonic coherent states mentioned in reference \cite{Wang1}.
Furthermore, the following bosonic and fermionic coherent states
\begin {equation}
|{\psi'}\rangle_{boson}=\frac{1}{\sqrt{2}}|\alpha\rangle_+|{\alpha}\rangle_-+|{\alpha}\rangle_-|{\alpha}\rangle_+,
\end {equation}
\begin {equation}
|{\psi'}\rangle_{fermion}=
\frac{1}{\sqrt{2}}|{\theta}\rangle_+|{\theta}\rangle_{-}+|{\theta}\rangle_-|{\theta}\rangle_+,
\end {equation}
have the same form and both are MES, in the sense that the
$|{\psi'}\rangle_{boson}$ is MES by itself \cite{Wang3} and
\begin {equation}
\int d\theta^* d\theta\ (\frac{\theta^*}{4})\
|{\psi'}\rangle_{fermion}=|{\Psi}^{+}\rangle,
\end {equation}
which is clearly maximal entangled state. There are some other FCSs
that lead to MESs for fermionic coherent systems in integration
method which also have the maximally entangled bosonic counterpart
obtained by concurrence. For example, in the case $1,2$, we take the
plus case and $k_{1}=3,k_{2}=-1,k_{3}=1,k_{4}=-3$, then the state
(\ref{generalfer}) reduces to
$$
|3\theta\rangle|-\theta\rangle-|\theta\rangle|-3\theta\rangle,
$$
which its bosonic counterpart
$$
|3\alpha\rangle|-\alpha\rangle-|\alpha\rangle|-3\alpha\rangle.
$$
is also MES \cite{Wang1,Wang3}. Thus, according to (\ref{condfer}),
allocating an arbitrary value to $m$ and accounting for proper
conditions among $ k_{1},k_{2},k_{3},k_{4}$, one can obtain other
MESs for FCSs. As another example, let $m=3$, $ k_{1}=k_{2}=1,$ and
$k_{3}=k_{4}=-2$, then the state (\ref{generalfer}) reduces to
$$
|\theta\rangle|\theta\rangle-|-2\theta\rangle|-2\theta\rangle.
$$
which just like the above cases has the maximally entangled bosonic
counterpart. It is perhaps worth pointing out that, although it is
possible to find maximal FCSs (which have the same form in the
bosonic maximal coherent state of the form (\ref{bospn genera2})),
it can be shown that the inverse does not hold. For instance, we
have
$$
|{\Psi}\rangle_{max}=\int d\theta^* d\theta
\left(\frac{\theta^*}{\sqrt{2}(i-
1)}\right)[|i\theta\rangle|i\theta\rangle-|\theta\rangle|\theta\rangle]
$$
\begin {equation}
\hspace{1cm}= \frac{1}{\sqrt{2}}(|01\rangle+|10\rangle)=
|{\Psi}^{+}\rangle,
\end {equation}
which is clearly maximal while its bosonic counterpart
$$
|i\alpha\rangle|i\alpha\rangle-|\alpha\rangle|\alpha\rangle,
$$
is not. This is due to the fact that although
$Re(f_{13})=Re(f_{24}^*)$, we have $Im(f_{13})\neq Im(f_{24}^*)$
which implies that $f_{13}\neq f_{24}^*$, so the condition
(\ref{imagecon}) is not satisfied. In fact, the $k_{i}$s that
satisfy (\ref{realcon}) and (\ref{imagecon}) for bosonic coherent
states (\ref{bospn genera2}), must also satisfy the relaxed
conditions (\ref{condfer}) for FCSs (\ref{generalfer}).
\subsection{Example 2}
Now consider the following state
\begin {equation}\label{bospn genera3}
|k_{1}\alpha\rangle|k_{2}\alpha\rangle+|k_{3}\alpha\rangle|k_{4}\alpha\rangle\quad,
\qquad k_{i}\in \mathbb{C},
\end {equation}
which has  concurrence
\begin {equation}
C=\frac{2\left[(1-e^{{-\frac{1}{2}|\alpha|^2}(f_{13}+f_{13}^*)})(1-e^{{-\frac{1}{2}|\alpha|^2}(f_{24}+f_{24}^*)})\right]^{1/2}}
{2+e^{{-\frac{1}{2}|\alpha|^2(f_{13}^*+f_{24}^*)}}+e^{{-\frac{1}{2}|\alpha|^2(f_{13}+f_{24})}}},
\end {equation}
thus, the state (\ref{bospn genera3}) is maximally entangled when $
f_{13}=f_{24}^*+\frac{2i\pi}{|\alpha|^2}, $ in other word
\begin {equation}\label{realcon p}
|k_{1}-k_{3}|=|k_{2}-k_{4}|\quad,\ \ \mathrm{or} \quad
(k_{1}-k_{3})=(k_{2}-k_{4})e^{i\phi}
\end {equation}
and
\begin {equation}\label{imagecon p}
Im(k_4^*k_2)-Im(k_1^*k_3)=\frac{\pi}{|\alpha^2|}.
\end {equation}
Now let us consider the  same state as (\ref{bospn genera3}) but
complex $\alpha$ is replaced by Grassmann number $\theta$, i.e.,
\begin {equation}\label{generalfer2}
|k_{1}\theta\rangle|k_{2}\theta\rangle+|k_{3}\theta\rangle|k_{4}\theta\rangle.
\end {equation}
Again we take $ w(\theta,\theta^*)=\frac{1}{m\sqrt{2}}\theta^*, $
then
$$
|\Psi'\rangle_{max}=\int d\theta^* d\theta \
(\frac{\theta^*}{m\sqrt{2}})[|k_{1}\theta\rangle|k_{2}\theta\rangle+|k_{3}\theta\rangle|k_{4}\theta\rangle]
$$
\begin {equation}\label{int-gen2}
\hspace{1cm}= \frac{1}{m\sqrt{2}}[k_{2}+ k_{4}]|01\rangle+
\frac{1}{m\sqrt{2}}[k_{1}+ k_{3}]|10\rangle.
\end {equation}
This state is MES if
\begin {equation}\label{condfer2}
(k_{1}+k_{3})=e^{i\phi}(k_{2}+k_{4})=m.
\end {equation}
Here we can treat three cases separately.
\\
\textbf{Case 1:}\quad
 In the first case, let $k_{i}s $ satisfy
all the conditions (\ref{realcon p}), (\ref{imagecon p}) and
(\ref{condfer2}). Hence there is a fermionic counterpart for any
bosonic MES and vis versa. For example if $
k_{1}=k_{2}=\frac{i\pi}{2|\alpha|^2} $ , $ k_{3}=k_{4}=1$ and
$\phi=0$, then the Eq.(\ref{int-gen2}) gives MES for following FCS
\begin {equation}
|\frac{i\pi}{2|\alpha|^{2}}\theta\rangle|\frac{i\pi}{2|\alpha|^{2}}\theta\rangle+|\theta\rangle|\theta\rangle,
\end {equation}
and the state (\ref{bospn genera3}) reduces to the following MES for
bosonic coherent state
\begin {equation}
|\frac{i\pi}{2|\alpha|^{2}}\alpha\rangle|\frac{i\pi}{2|\alpha|^{2}}\alpha\rangle+|\alpha\rangle|\alpha\rangle,
\end {equation}
These states are counterparts of each other.
\\
\textbf{Case 2:}\quad In the second case, let $k_{i}s $ satisfy the
conditions (\ref{realcon p}), (\ref{imagecon p}) but the condition
(\ref{condfer2}) does no hold. Therefor we have a set of bosonic
MESs which  have no similar fermionic maximally entangled
counterparts. For example
\begin {equation}
|(\frac{\pi}{2|\alpha|^{2}}+i)\alpha\rangle|(\frac{\pi}{|\alpha|^{2}}+i)\alpha\rangle+
|(\frac{\pi}{2|\alpha|^{2}}-i)\alpha\rangle|(\frac{\pi}{|\alpha|^{2}}-i)\alpha\rangle.
\end {equation}
Clearly the fermonic counter part of this state does not lead to a
MES with any choice of weight function.
\\
\textbf{Case 3:}\quad In the third case, let $k_{i}s $ satisfy the
conditions (\ref{condfer2}) but  the condition (\ref{realcon p}) or
(\ref{imagecon p}) does no hold. Hence we have a set of fermionic
coherent states that, according to Eq.(\ref{int-gen2}),  give MESs
while the bosonic counterpart of them are not MESs. To give an
example  we can take FCS
\begin {equation}\label{generalfer2}
|k\theta\rangle|l\theta\rangle+|l\theta\rangle|k\theta\rangle\quad,
\qquad k , l\in \mathbb{C},
\end {equation}
which have no maximally entangled bosonic counterpart.
\section{Conclusion}
In summary, we have shown the some well-known  entangled pure states
like {\bf{GHZ}}, {\bf{W}}, Bell, Bell-like and biseparable states
can be constructed by tensor product of fermion coherent states with
integration over proper Grassmann weight functions. For three qubit
{\bf{GHZ}} and  {\bf{W}} states, the construction can be easily
generalized to multi-qubit cases, however there is an important
difference between {\bf{GHZ}} and {\bf{W}} constructions: in  the
former case, we must use tensor product of FCSs with different
Grassmann numbers, while in the latter case the tensor product of
$n$ FCSs $|\theta\rangle$ is sufficient to this aim. We called a FCS
\emph{maximal}, if there is a Grassmann weight function whose
integration over that FCS gives a MES. As we saw in the last
section, some maximally entangled BCSs have FCSs counterparts, but
it is of course perfectly possible to find simple examples of
maximal FCSs, using the integration method, which have no  maximal
BCSs counterparts and vice versa.

\end{document}